\documentclass[journal,twoside,web]{ieeecolor}
\usepackage{generic}
\usepackage{amsmath,dsfont,amssymb}

\usepackage{amsthm}
\usepackage{mathtools}
\usepackage{graphicx}
\usepackage{amsfonts}
\usepackage{float}
\usepackage{mathrsfs}
\usepackage{hybridsystem}
\DeclareMathOperator{\su}{\text{Succ}}

\DeclareMathOperator{\bfz}{\mathbf{z}}
\DeclareMathOperator{\bfu}{\mathbf{u}}
\DeclareMathOperator{\bfx}{\mathbf{x}}
\DeclareMathOperator{\pre}{\text{Pre}}
\DeclareMathOperator{\ch}{\text{Ch}}

\usepackage[noadjust]{cite} 

\usepackage[hidelinks]{hyperref}
\usepackage{algorithmicx}
\usepackage{algorithm}
\usepackage{algpseudocode}
\usepackage{lipsum}
\usepackage{subfig}
\usepackage{amssymb}
\usepackage{bbm}
\usepackage{float}
\usepackage{balance}
\usepackage{color}
\theoremstyle{plain}
\newtheorem{thm}{\textbf{Theorem}}

\theoremstyle{definition}
\newtheorem{defn}{\textbf{Definition}}

\theoremstyle{remark}
\newtheorem{rem}{\textbf{Remark}}

\allowdisplaybreaks
\usepackage{algpseudocode}
\algdef{SE}[DOWHILE]{Do}{DoWhile}{\algorithmicdo}[1]{\algorithmicwhile\ #1}%

\DeclareMathOperator*{\esssup}{ess\,sup}

\usepackage[normalem]{ulem}

\usepackage{hyperref}
\hypersetup{colorlinks=true,
linkcolor = blue,
urlcolor=cyan,
citecolor =red
}
\begin{document}

\title{\LARGE \bf
Interactive multi-modal motion planning with Branch Model Predictive Control
}
\author{Yuxiao Chen, Ugo Rosolia, Wyatt Ubellacker, Noel Csomay-Shanklin, and Aaron D. Ames
\thanks{The authors are with the Department of Mechanical and Civil Engineering, California Institute of Technology, Pasadena, CA, USA.{chenyx,urosolia,wubellac,noelcs,ames}@caltech.edu}
}

\maketitle
\thispagestyle{empty}

\begin{abstract}
Motion planning for autonomous robots and vehicles in presence of uncontrolled agents remains a challenging problem as the reactive behaviors of the uncontrolled agents must be considered. Since the uncontrolled agents usually demonstrate multimodal reactive behavior, the motion planner needs to solve a continuous motion planning problem under these behaviors, which contains a discrete element. We propose a branch Model Predictive Control (MPC) framework that plans over feedback policies to leverage the reactive behavior of the uncontrolled agent. In particular, a scenario tree is constructed from a finite set of policies of the uncontrolled agent, and the branch MPC solves for a feedback policy in the form of a trajectory tree, which shares the same topology as the scenario tree. Moreover, coherent risk measures such as the Conditional Value at Risk (CVaR) are used as a tuning knob to adjust the tradeoff between performance and robustness. The proposed branch MPC framework is tested on an \textit{overtake and lane change} task and a \textit{merging} task for autonomous vehicles in simulation, and on the motion planning of an autonomous quadruped robot alongside an uncontrolled quadruped in experiments. The result demonstrates interesting human-like behaviors, achieving a balance between safety and performance.
\end{abstract}

\section{Introduction}\label{sec:intro}
Motion planning is one of the central modules of autonomous robots and vehicles. In particular, as the autonomous agent shares the environment with uncontrolled agents, the reactive behaviors of the uncontrolled agents need to be taken into account in the motion planning. Take autonomous vehicles as an example: as they share the road with uncontrolled vehicles, pedestrians, and cyclists, and all of whom would adjust their behaviors based on other agents around them, the motion planner for the autonomous vehicle then needs to plan trajectories that are safe yet not overly conservative in presence of other road users. The challenges of achieving safe interactive motion planning in presence of uncontrolled agents are twofold: (i) the reactive behavior of the uncontrolled agents need to be properly modelled (ii) the reactive behaviors need to be leveraged in the motion planning.

The modelling of reactive behaviors is a challenging problem that has attracted increasing attention recently. Many uncontrolled agents encountered by autonomous agents exhibit highly nondeterministic and multimodal behaviors. For instance, a human driver may choose totally different behaviors under the same situation at different times (swerve left or right, yield or not yield). Therefore, accurately modeling these nondeterministic behaviors is almost impossible. Stochastic models, naturally, have been proposed to model the nondeterministic reactive behavior, such as Markovian models \cite{althoff2009model,kumagai2006prediction,bai2015intention}, and generative models \cite{ivanovic2019trajectron,gupta2018social}. Another class of approaches is the set-based method that models the set of possible behaviors, including the GAN-based prediction \cite{bhattacharyya2018multi}, and classifier-based approaches \cite{phan2020covernet,chen2020reactive,chen2018modelling,chen2020counter}.

Once a predictive model is obtained, in most cases, the motion planner is at the downstream of the predictive model. The typical strategy is to use some form of robust motion planning algorithm to let the autonomous agent avoid the reachable set of the uncontrolled agent \cite{kousik2017safe} or the set of all possible trajectories \cite{chen2020reactive}. Obviously, the robust formulations lead to conservative motion plans and compromised performance. The main source of conservatism of robust planning is that it ignores the reactivity of the uncontrolled agent within the prediction horizon. In particular, a typical predictive model takes the scenario description as input and outputs the prediction of the uncontrolled agent's behavior, and these behaviors are assumed to be fixed by the motion planner under the robust planning setup. This is not a deliberate choice, but rather a compromise due to the difficulty of reasoning about future reactive behaviors. Since the future scenario is nondeterministic as the behavior of the uncontrolled agent is nondeterministic, the reasoning of the behavior of the uncontrolled agent in future prediction steps is usually convoluted and intractable.

In fact, the tractability of the reactive motion planning problem heavily depends on the problem description. To be specific, when the action space of the uncontrolled agent is discrete and finite, such as in Markov decision processes (MDPs) and Partially Observed Markov Decision Processes (POMDPs), future scenarios can be enumerated and the planning can be solved with dynamic programming (DP) \cite{kurniawati2008sarsop}. The formulation of MDPs is naturally multimodal as each outcome of the state transition can be viewed as a mode. However, MDPs with discrete action and state spaces are too coarse for many motion planning problems and cannot generate trajectories practical for highly dynamic systems. We note that MDPs have been extended to systems with continuous input and state spaces \cite{brooks2006parametric,patil2015scaling}, typically under simplifying assumptions such as Gaussian observation and process noise, and maximum likelihood observation. However, these simplifications usually cause the solution to lose multimodality.

To account for the reactive behaviors in future steps, we propose a branching Model Predictive Control (MPC) framework that combines the continuous motion planning with discrete modes representing the multimodal reactive behaviors of the uncontrolled agent. To obtain tractability, the continuous spectrum of possible behaviors of the uncontrolled agent is simplified with a finite set of policies, which are feedback control laws that represent the multimodality of the reactive behavior. On top of this finite set of policies, we assume that a predictive model is given that outputs the probability of each policy given the scenario. We then construct a scenario tree by forward propagating the policies of the uncontrolled agent and solve for a trajectory tree with the same topology. The objective is then to minimize the expected overall cost, where the expectation is taken over all the branches with the probability given by the predictive model. Furthermore, we use coherent risk measures such as the Conditional Value at Risk (CVaR) in presence of the possibly inaccurate predictive model to improve the robustness of the policy. The branching idea for MPC can date back to \cite{scokaert1998min}. In~\cite{scokaert1998min}, the authors showed that optimizing over a tree of trajectories is equivalent to optimizing over feedback policies, when the goal is to minimize the worst case cost and the uncertainties acting on the system are uni-modal.  

The multi-modal case was studied in~\cite{sopasakis2019risk, alsterda2021contingency, alsterda2019contingency, batkovic2020robust}, where the authors leveraged the optimization over a tree of trajectories to optimize over a set of feedback policies. However in these works the interaction between the controlled agent and the environment was not considered, meaning that the probability associated with the uncertainty modes determining the different tree branches is fixed. 
The advantage of the proposed branch MPC framework are (i) the reactive behavior of the uncontrolled agent in future steps is leveraged via the branching structure (ii) the MPC formulation can handle highly dynamic systems and generate high-quality motion plans (iii) the risk level acts as a convenient tuning knob reflecting the confidence of the reactive model and enables tradeoff between robustness and performance.

\section{Preliminaries}
We first review the major tools used in our framework.
\subsection{Model predictive control}
Model Predictive Control (MPC) is a well-established control methodology that leverages forecast to compute actions~\cite{borrelli2017predictive, kouvaritakis2016model}. In MPC, at each time $t$ the controller plans a sequence of open-loop actions to minimize an objective function, and then the first predicted action is applied to the system. More formally, at each time $t$ given the state of the system $x(t)$ the action is computed solving the following \textit{Finite Time Optimal Control Problem (FTOCP)}:
\begin{equation*}
\begin{aligned}
    \min_{u_0,\ldots, u_{N-1}} \quad & \sum_{k = 0}^{N-1} h(x_k, u_k) + V(x_N) \\
    \text{s.t. } \quad\quad & x_{k+1} = f(x_k, u_k), \forall k \in \{0, \ldots, N-1\},\\
    & x_k \in \mathcal{X}, u_k \in \mathcal{U},  \forall k \in \{0, \ldots, N-1\}, \\
    & x_0 = x(t), x_N \in \mathcal{X}_N,
\end{aligned}
\end{equation*}
where the discrete time update $x_{k+1} = f(x_k, u_k)$ describes the evolution of the state $x_k \in \mathbb{R}^n$ when the input $u_k \in \mathbb{R}^d$ is applied to the system. In the above FTOCP, the stage cost $h:\mathbb{R}^n \times\mathbb{R}^d \rightarrow \mathbb{R}$ and the terminal cost $V : \mathbb{R}^n \rightarrow \mathbb{R}$ are defined by the control designer. Furthermore, $\mathcal{X} \subset \mathbb{R}^n$, $\mathcal{U} \subset \mathbb{R}^d$, and $\mathcal{X}_N \subset \mathbb{R}^n$ are the state, input, and terminal constraint sets, respectively. At time $t$, let \begin{equation*}
    [u_0^*(x(t)), \ldots, u_{N-1}^*(x(t))]
\end{equation*}
be the optimal input sequence to the above FTOCP, which is a function of the current systems's state $x(t)$. Then, the first optimal action is applied, resulting in the following MPC policy:
\begin{equation*}
    \pi^{\text{MPC}}(x(t)) = u_0^*(x(t)).
\end{equation*}
\subsection{Risk-aware decision making}\label{sec:risk_review}
Risk-aware decision making concerns an optimization problem with stochastic outcomes \cite{ruszczynski2006optimization}. Rather than optimizing over the expectation of the cost function, risk-aware optimization optimizes over a risk measure, which is a functional of the probability distributions over the cost function. Formally, consider a probability space $(\Omega,\mathscr{F},P)$ where $\Omega$ is a Borel measurable space with a $\sigma$-algebra $\mathscr{F}$ and probability function $P$. The cost function then can be understood as a function $X:\Omega\to\mathbb{R}$. We let $\mathscr{X}$ denote the linear space of all $\mathscr{F}$-measurable functions. A risk measure $\rho(\cdot)$ is a functional that maps from $\mathscr{X}$ to the extended real line $\{-\infty\}\cup\mathbb{R}\cup\{+\infty\}$. In particular, one useful class of risk measures are the \textit{coherent risk measures}, defined below.
\begin{defn}
  A risk measure $\rho$ is a \textit{coherent risk measure} if it satisfies the following properties
  \begin{itemize}
    \item \textbf{Convexity}: $\forall X,Y\in \mathscr{X},\forall \lambda\in[0,1], \rho(\lambda X+(1-\lambda)Y)\le \lambda \rho(X)+(1-\lambda)\rho(Y)$
    \item \textbf{Monotonicity}: $\forall X,Y\in\mathscr{X}$ with $Y\ge X$, $\rho(Y)\ge \rho(X)$.
    \item \textbf{Translation equivariance}: For all $c\in\mathbb{R}$ and $X\in\mathscr{X}$, $\rho(X+c)=\rho(X)+c$
    \item \textbf{Positive homogeneity}: For all $t>0$, $X\in\mathscr{X}$, $\rho(tX)=t\rho(X)$.
  \end{itemize}
\end{defn}
As a counterexample, value at risk (VaR), which is defined as $\text{VaR}_{1-\alpha}(X):=\min_{z\in\mathbb{R}:P(X\ge z)\le \alpha} z$ is not a coherent risk measure as it does not satisfy the convexity property. Examples of coherent risk measures include the conditional value at risk (CVaR), entropic value at risk (EVaR) \cite{ahmadi2012entropic,dixit2020risk}, and expectation.
\begin{defn}
For a random variable $X:\Omega\to\mathbb{R}$, the conditional value at risk is defined as
\begin{equation}\label{eq:CVaR}
\begin{aligned}
  \text{CVaR}_{1-\alpha}(X)&:=\frac{1}{\alpha}\int_0^\alpha \text{VaR}_{1-\gamma}(X)d \gamma\\
  &=\inf_{z\in\mathbb{R}}\{z+\frac{1}{\alpha}\int_{-\infty}^\infty [x-z]_+dP(x)\},
  \end{aligned}
\end{equation}
where $dP(x)$ is the probability density function of $X$.
\end{defn}
CVaR is the expectation of the $\alpha$ percent worst outcomes of $X$, $\alpha\in[0,1]$ is a parameter that determines how risk-averse the CVaR is and CVaR is monotonically decreasing w.r.t. $\alpha$. Moreover, $\lim_{\alpha\to 0}\text{CVaR}_{1-\alpha}(X)=\esssup X$ and $\text{CVaR}_{0}(X)=\mathbb{E}[X]$.

It is shown in~\cite{artzner1999coherent,rockafellar2002deviation} that any coherent risk measure has a dual representation as the maximization of expectation over a probability ambiguity set, which often turns out more convenient in computation, i.e.,
\begin{equation*}
  \rho(X)=\max_{Q\in\mathscr{A}}\mathbb{E}_Q [X].
\end{equation*}
$\mathscr{A}$ is the ambiguity set, which is a set of probability distributions. For CVaR, the ambiguity set is
\begin{equation}\label{eq:CVAR_amb}
  \mathscr{A}=\left\{Q\left|\begin{aligned}&\forall x\in\mathbb{R}, dQ(x)\ge 0,\\&\int dQ(x)dx=1, dQ(x)\le \frac{1}{\alpha}dP(x)\end{aligned} \right.\right\}.
\end{equation}
When the distribution is discrete, the ambiguity set reduces to $\mathscr{A}=\{Q|Q(x_i)\ge0, \sum Q(x_i)=1, Q(x_i)\le \frac{1}{\alpha}P(x_i)\}$.
Risk-aware decision making then looks for the decision variable that minimizes some risk measure. The typical solution is via dualization, which we present in detail in Section \ref{sec:risk}.

\section{Branch MPC}\label{sec:MPC}
To fully utilize the reactivity of the uncontrolled agent, the control strategy needs to adapt to the behavior of the uncontrolled agent and reason about its future reaction to the environment. The classic model predictive control formulation optimizes over a single trajectory with a finite horizon, which is not able to leverage the reactivity. \cite{scokaert1998min} proposed a min-max model predictive control which, instead of optimizing over a single trajectory, searches for a policy that reacts to different disturbance signals via enumerating the extreme cases of the disturbance signal. The enumeration generates a branching tree structure that grows exponentially with the planning horizon, and a min-max problem is solved to obtain the robust MPC policy. The branching enumeration bears some similarity to the forward dynamic programming method used in many decision-making problems such as the Markov decision processes and Partially observed Markov decision processes. The difference is that in Markov models, a transition probability is associated with each branch, and the goal is the expectation of the reward instead of the worst-case performance.

The branch MPC proposed in this paper extends the branch enumeration strategy proposed in \cite{scokaert1998min} and associates it with a probabilistic characterization of the branches via a predictive model. To be specific, a finite set of policies are propagated forward to generate a scenario tree representing possible future behaviors of the uncontrolled agent. The branch MPC then optimizes over feedback policies in the form of a trajectory tree, which shares the same topology as the scenario tree. Each branch in the trajectory tree is the instantiation of the feedback policy under the uncontrolled agent's behavior characterized by the corresponding branch in the scenario tree.

We consider the scenario with one ego agent under control and one uncontrolled agent. A scenario tree is constructed by enumerating the behavior of the uncontrolled agent, and the branching probability is a function of the state of the ego agent and the uncontrolled agent.
\begin{rem}\label{rem:causality}
When there are multiple uncontrolled agents, one needs to consider the product space of their behaviors. Due to the exponential complexity, some pruning protocol is needed, which is beyond the scope of this paper. We plan to tackle the case with multiple uncontrolled agents in future works. 
\end{rem}

We let $x\in\mathcal{X}\subseteq \mathbb{R}^{n_x}$ denote the state of the ego agent and $z\in\mathcal{Z}\subseteq \mathbb{R}^{n_z}$ denote the state of the uncontrolled agent, following known dynamics:
\begin{equation}\label{eq:dyn}
x^+=f(x,u),\quad z^+=g(z,d),
\end{equation}
where $u\in\mathcal{U}$ is the ego agent input, $d\in\mathcal{D}$ is the uncontrolled agent input.

\newsec{Policies for the uncontrolled agent.}
In order to enable branching within the prediction horizon, the behavior of the uncontrolled agent needs to be enumerable. However, using discrete actions, even motion primitives, can cause the behavior of the uncontrolled agent to be stiff and unnatural. Motivated by the fact that human driving behaviors demonstrate strong multimodality, a finite set of policies $\Pi=\{\pi_i\}_{i=1}^m$ is used to represent the possible behaviors of the uncontrolled agent, which consists of feedback policies $\pi_i:\mathcal{Z}\to\mathcal{D}$. Take highway driving as an example: the set of policies can include \textit{maintain fixed speed}, \textit{slow down}, \textit{left lane change} and \textit{right lane change}. Note that the feedback policies only depend on $z$, making it possible to be propagated forward independently of other agents in the scene.

\newsec{Predictive model and scenario tree.}
Given a policy set, $z$ can be propagated forward with a selected policy. A scenario tree is constructed by enumerating the policies. To be specific, a scenario tree starts at the root, which is the current state of the uncontrolled agent $z$ and propagates forward. Certain nodes in the tree are selected as branching nodes, which would have $m$ children, each following one of the policies in $\Pi$. A non-branching node only has one child, following its current policy. In particular, the root is always a branching node, and we use a simple strategy where branching happens every $M$ steps. Ideally, one wants all the nodes in the scenario tree to be branching nodes. However, this naive choice may lead to an unnecessarily large number of nodes, causing computation issues. A subset of consecutive nodes following the same policy is called a branch. A branch in the scenario tree ends at a branching node or a leaf node (nodes with no children) and contains all its non-branching predecessors up to its first predecessor that is a branching node.

\begin{figure}
  \centering
  \includegraphics[width=1\columnwidth]{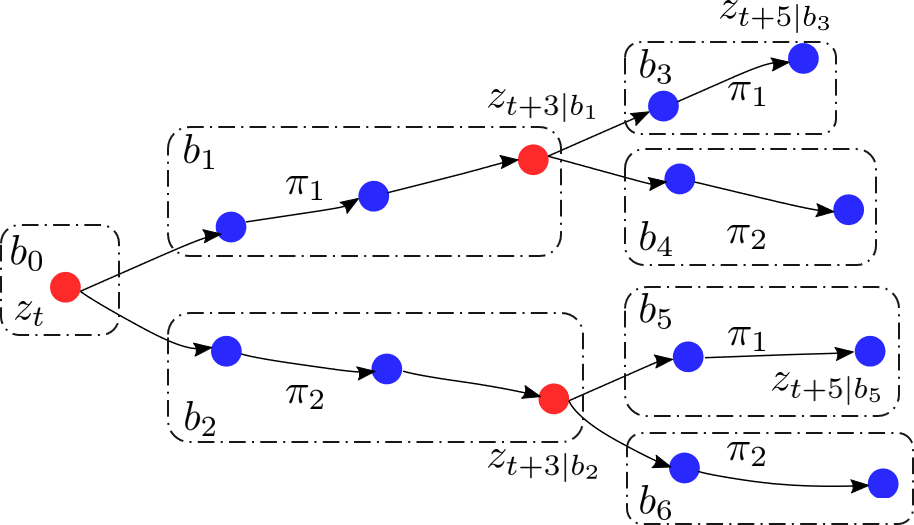}
  \includegraphics[width=1\columnwidth]{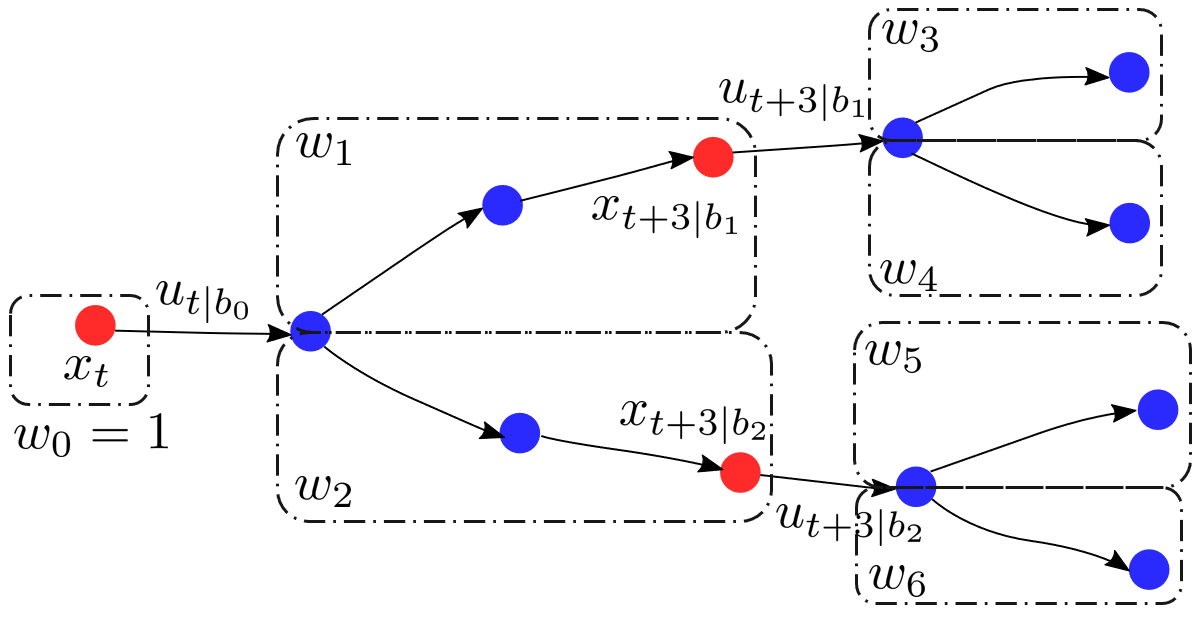}
  \caption{Example scenario tree with two policies, branching nodes are in red and non-branching nodes are in blue}\label{fig:scenario_tree}
\end{figure}

The top figure of Fig. \ref{fig:scenario_tree} shows an example of a scenario tree with $m=2, M=3$ (each branching node has 2 children, branching happens every 3 time steps), the dashed squares show the branches in the tree. The branch MPC would construct another tree, denoted as the trajectory tree, of the ego agent trajectory whose topology is similar to the scenario tree, each node contains the planned future state of the ego agent, which are variables of the MPC FTOCP, as shown in the bottom figure of Fig. \ref{fig:scenario_tree}. $b_0$ is the root branch with children $b_1$ and $b_2$, and $b_1$ has two children, $b_3$ and $b_4$, and so on. $\mathbf{I}$ denotes the set of indices of all the branches in the scenario tree. For the one in Figure \ref{fig:scenario_tree}, $\mathbf{I}=\{0,1,2,3,4,5,6\}$.
\begin{rem}
Due to the causality constraint, before the uncontrolled agent demonstrate which policy it executes, the MPC would not be able to take different actions. Therefore, the node directly following a branching node in the trajectory tree is shared by multiple branches. Another way to view it is that the $m$ children of a branching node share the same state in the trajectory tree.
\end{rem}
For a branch $b_i$, $w_i$ denotes its weight, the root always has weight $w_0=1$, and at every branching point, a predictive model then assigns weight to the children of the branching node based on the $x$ and $z$ predictions. Take the branching from $b_1$ to $b_3$ and $b_4$ as an example,
\begin{equation}\label{eq:predictive_model}
\begin{aligned}
w_3&=w_1 P(\pi_1|x_{t+3|b_1},z_{t+1|b_1}),\\
w_4&=w_1 P(\pi_2|x_{t+3|b_1},z_{t+1|b_1}),
\end{aligned}
\end{equation}
where $x_{t+3|b_1}$ is the state prediction at $t+3$ in branch $b_1$, $P(\pi|x,z)$ is the probability of the uncontrolled agent taking $\pi$ under the scenario described by $x$ and $z$, which is given by the predictive model, and $\sum_{i=1}^m P(\pi_i,x,z)=1$ always holds. An evaluation of the scenario tree refers to a path from the root to one of the leaf nodes. For example, $b_0\to b_1 \to b_3$ is an evaluation of the scenario tree shown in Fig. \ref{fig:scenario_tree}.

\newsec{Branch MPC setup.}
For notational clarity, we let $t^0_i$ and $t^f_i$ denote the time instances of the first and last node in $b_i$, and let $\bfx_i=[x_{t_i^0|b_i},...,x_{t_i^f|b_i}]$, $\bfz_i=[z_{t_i^0|b_i},...,z_{t_i^f|b_i}]$, $\bfu_i=[u_{t_i^0|b_i},...,u_{t_i^f|b_i}]$ be the ego agent trajectory, uncontrolled agent trajectory, and ego agent control input sequence of $b_i$. $\pre(\cdot)$ denotes the parent of a branch and $\ch(\cdot)$ denote the set of children of a branch, e.g., $\pre(b_1)=b_0$, $\ch(b_1)=\{b_3,b_4\}$. Sometimes with a slight abuse of notation, we use $\ch(\cdot)$ to denote the set of indices of the children branches. When the context is clear, we also use $P(b_i|\bfx_{\pre(i)},\bfz_{\pre(i)})$ to denote the branching probability of $b_i$ given the $\bfx$ and $\bfz$ of its parent branch. For each branch in the scenario tree, a cost function $J_i(\bfx_i,\bfz_i,\bfu_i)$ and a set of constraints $\mathcal{C}_i(\bfx_i,\bfz_i,\bfu_i)\le 0$ are considered. We use slack variables to define the extended local cost function:
\begin{equation*}
  \bar{J}_i(\bfx_i,\bfz_i,\bfu_i)=J_i(\bfx_i,\bfz_i,\bfu_i)+\beta\mathds{1}^\intercal[\mathcal{C}_i(\bfx_i,\bfz_i,\bfu_i)]_+,
\end{equation*}
where $[\cdot]_+=\max\{0,\cdot\}$, $\beta$ is the slack penalty.
Given the ego agent state $x_t$ and the predicted trajectory $\bfz$ for all $i\in \mathbf{I}$, the overall branch MPC then solves the following optimization problem:
\begin{subequations}\label{eq:branch_MPC}
\begin{align}
\min_{\{\bfx_i,\bfu_i\}_\mathbf{I}}~&\sum_{i\in\mathbf{I}} w_i(\bfx,\bfz)\bar{J}_i(\bfx_i,\bfz_i,\bfu_i) \nonumber\\
\mathrm{s.t.}~~&  x_{t|b_i}=f(x_{t-1|b_i},u_{t-1|b_i}), \forall i\in \mathbf{I},\;\forall t=t_i^0+1,...t_i^f, \label{eq:inbranch}\\
&x_{t_k^0|b_k}=f(x_{t_k^0-1|\pre(b_k)},u_{t_k^0-1|\pre(b_k)}), \forall k\in\mathbf{I}\backslash \{0\}\label{eq:connect},\\
& w_0=1,w_i=w_{Pre(i)}P(b_i|\bfx_{\pre(i)},\bfz_{\pre(i)})\label{eq:weight_evo}\\
& x_{t_0^0}=x_t,z_{t_0^0}=z_t,\label{eq:init_cond}
\end{align}
\end{subequations}
where \eqref{eq:inbranch} enforces the dynamics within the branches, \eqref{eq:connect} enforces the dynamics on the connection between any branch and its parent. We use $w_i(\bfx,\bfz)$ to indicate that the branch weight depends on the planned trajectory $\bfx$ and the predicted $\bfz$. \eqref{eq:weight_evo} encodes the weights of the branches with the predictive model, and \eqref{eq:init_cond} is the constraint on the initial condition. As previously noted in Remark \ref{rem:causality}, for causality, since all children share the same input at every branching point, the first node of all children branches must share the same state.

The optimization in \eqref{eq:branch_MPC} tries to minimize the expected total cost over the prediction horizon, which is a weighted sum of the cost over the branches in the scenario tree. In general, \eqref{eq:branch_MPC} is nonconvex, and we use the Sequential Quadratic Program (SQP) approach to solve it by linearizing the dynamics, the constraints, and the weights of the branches. To be specific, at every iteration, the solution from the last iteration is used as the linearization point. The dynamics after linearization is simply
\begin{equation*}
  x_{t+1|b_i}=A_{t|b_i}x_{t|b_i}+B_{t|b_i}u_{t|b_i}+C_{t|b_i},
\end{equation*}
where $A_{t|b_i}=\frac{\partial f(x,u)}{\partial x}|_{\hat{x}_{t|b_i},\hat{u}_{t|b_i}}$, $B_{t|b_i}=\frac{\partial f(x,u)}{\partial u}|_{\hat{x}_{t|b_i},\hat{u}_{t|b_i}}$, $C_{t|b_i}=f(\hat{x}_{t|b_i},\hat{u}_{t|b_i})-A_{t|b_i}\hat{x}_{t|b_i}-B_{t|b_i}\hat{u}_{t|b_i}$. State and input values shifted backwards from the last iteration are given by $\hat{x}_{t|b_i},\hat{u}_{t|b_i}$, i.e., $\hat{x}_{t|b_i}$ is the solution of the state associated with its children in the scenario tree from the last iteration. The constraints are linearized and enforced as linear inequality constraints (with slack). The cost function follows a Linear Quadratic Regulator (LQR) type format as $(x-x_{ref})^\intercal Q (x-x_{ref})+u^\intercal R u$ with PSD matrices $Q$ and $R$. Although the quadratic cost does not need not to be linearized, the product of the cost and the weight $w_i$ on every branch needs to be linearized:
\begin{equation}\label{eq:linearize_w}
\begin{aligned}
 w_i(\bfx_i,\bfz_i) \bar{J}_i(\bfx_i,\bfz_i,\bfu_i)&\approx \frac{\partial w_i}{\partial \bfx}|_{\hat{\bfx}_i} \bar{J}_i(\hat{\bfx}_i,\bfz_i,\hat{\bfu}_i)\\
 &+ w_i(\hat{\bfx}_i,\bfz_i) \bar{J}_i(\bfx_i,\bfz_i,\bfu_i),
 \end{aligned}
\end{equation}
where $\hat{\bfx}_i$ and $\hat{\bfu}_i$ are the solution from the last iteration shifted backwards. The first term on the RHS of \eqref{eq:linearize_w} represents the change of the weighted cost function achieved by changing $w_i$, and the second term represents the cost with the weight calculated with the motion plan from the last iteration. This decomposition of cost encourages the ego agent to not only minimize the weighted cost by minimizing $\bar{J}$, but also put more weight on branches with smaller costs via influencing the weights given by the predictive model. This phenomenon is shown later in the vehicle highway simulation where the ego vehicle would `nudge' forward before an overtaking to reduce the probability that the uncontrolled vehicle changes lane and blocks the overtaking.

The SQP after linearization for branch MPC is then
\begin{equation}\label{eq:linearized_MPC}
  \begin{aligned}
\min_{\bfx_i,\bfu_i}\quad&\sum_{i\in\mathbf{I}} \frac{\partial w_i}{\partial \bfx}|_{\hat{\bfx}_i} \bar{J}_i(\hat{\bfx}_i,\bfz_i,\hat{\bfu}_i) + w_i(\hat{\bfx}_i,\bfz_i) \bar{J}_i(\bfx_i,\bfz_i,\bfu_i) \\
\mathrm{s.t.}\quad& x_{t|b_k}= A_{t-1|b_k}x_{t-1|b_k}+B_{t-1|b_k}u_{t-1|b_k}+C_{t-1|b_k}, \\
&\quad\quad\quad\quad\quad\quad\quad\quad\quad\quad\quad  \forall i\in \mathbf{I},\forall t=t_i^0+1,...t_i^f,\\
&x_{t_k^0|b_k}=A_{t'|b_j}x_{t'|b_j}+B_{t'|b_j}u_{t'|b_j}+C_{t'|b_j},\\
&\quad\quad\quad\quad\quad\quad\forall k\in\mathbf{I}\backslash \{0\}, b_j=\pre(b_k),t'=t_k^0-1,\\
\end{aligned}
\end{equation}
After each iteration, only the first input is used as control command and the MPC solver replans at every iteration. However, note that the root is the ancestor of the whole scenario tree and the first input influences all the subsequent branches, indicating that the branch MPC policy takes all the subsequent branches in the scenario tree into account and leverages the reactive behavior in future predictions steps.

\section{Branch MPC with risk measure objectives}\label{sec:risk}
The previous setup of branch MPC is essentially minimizing the expectation of the cost over the scenario tree, which is risk-neutral. Suppose one is not very confident about the predictive model, or one is more cautious, i.e., cares more about the bad outcome of the scenario tree; then, a risk-averse policy is preferred. By optimizing over a risk measure of the cost function, more focus is put on the worse outcomes. We use CVaR, which is the mean of the $\alpha$-percentile worst outcomes. The risk-averseness is conveniently tuned by the parameter $\alpha$, which is used as a tuning knob for the tradeoff between performance and robustness in our MPC setup.

As mentioned in Section \ref{sec:risk_review}, the dual representation of the risk measure gives another interpretation of the CVaR optimization. Take a discrete random variable $X$ as an example, and suppose there are $N$ outcomes $\xi_1,... \xi_N$ with probabilities $p_1,...p_N$ and $\sum p_i=1$. The expectation is $\mathbb{E}[X]=\sum_i p_i \xi_i$, and using the dual representation, the CVaR of $X$ is
\begin{equation*}
\begin{aligned}
\text{CVaR}_{1-\alpha} (X) &= \max_{P\in\mathscr{A}} \mathbb{E}_P[X],\\
\mathscr{A}&=\{q\in\mathbb{R}^N|q_i\ge 0,\sum_i^N q_i=1,q_i\le \frac{1}{\alpha}p_i\}.
\end{aligned}
\end{equation*}
The ambiguity set $\mathscr{A}$ can be viewed as a set of probability distributions similar to the assumed distribution of $X$, and $\alpha$ determines the maximum allowed difference. From this view point, the CVaR optimization is a robust optimization over distributions close to the one given by the predictive model.

The stochasticity of the branch MPC comes from the predictive model, which determines the weights of the branches. Since we work with a scenario tree with multiple stages of branching, we adopt the nested risk measure discussed in \cite{sopasakis2019risk,singh2018framework}. In particular, let $\phi_i$ denote the cost incurred by all the subsequent branches of $b_i$, which is a random variable whose value depends on the evaluation of the scenario tree, i.e., which policy the uncontrolled agent choose in reality. For example, for $b_0$ in Fig. \ref{fig:scenario_tree}, $\phi_0=\bar{J}_1+\bar{J}_3$ if the evaluation is $b_0\to b_1\to b_3$, and the probability associated with this evaluation is $w_3$. The risk measure is then defined in a nested way. Let $\rho_i$ be the risk measure of $\phi_i$ and let $\mathscr{A}_i$ denote the ambiguity set corresponding to the risk measure, then
\begin{equation}\label{eq:nested_risk}
    \rho_i(\phi_i)=\max_{Q\in \mathscr{A}_i} \mathbb{E}_Q[\bar{J}_{\su(i)}+\rho_{\su(i)}(\phi_{\su(i)})],
\end{equation}
where $\su(i)$ is the succeeding branch of $b_i$, which is a random variable, and $\mathscr{A}_i$ is a set of probability distributions over $\su(i)$, determined by the risk measure. Take the scenario tree in Fig. \ref{fig:scenario_tree} as an example: it contains 3 layers, each branching node has two possible successors. Then
\begin{equation}
\begin{aligned}
J &= \bar{J}_0+\rho_0(\phi_0) \\
\rho_0(\phi_0)& = \max_{Q\in\mathscr{A}_0} Q(b_1) (\bar{J}_1+\rho_1(\phi_1))+Q(b_2) (\bar{J}_2+\rho_2(\phi_2))\\
\rho_1(\phi_1)& = \max_{Q\in\mathscr{A}_1} Q(b_3) \bar{J}_3 +Q(b_4) \bar{J}_4\\
\rho_2(\phi_2)& = \max_{q\in\mathscr{A}_2} Q(b_5) \bar{J}_5 +Q(b_6) \bar{J}_6.
\end{aligned}
\end{equation}
If $\rho_i$ is taken to be $\text{CVaR}_\alpha(\cdot)$, the ambiguity set is simply $\mathscr{A}_i=\{Q|\forall j\in \ch(b_i),Q(b_j)\ge 0, \sum_{j\in\ch(i)} Q(b_j)=1,Q(b_j)\le \frac{1}{\alpha}P(b_j|\bfx_i,\bfz_i)\}$, where $P(b_j|\bfx_i,\bfz_i)$ is the branching probability at $b_i$. Note that $\rho_1$ and $\rho_2$ are nested in the definition of $\rho_0$.

Under the dual formulation, the optimization over risk measures is essentially minimizing the worst-case expectation over distributions in the ambiguity set, which is a distributionally robust optimization (DRO) \cite{rahimian2019distributionally}. It turns out that CVaR has a convenient conic representation \cite{sopasakis2019risk} and with Lagrange duality, one can enforce the nested risk measure as convex constraints via the epigraph representation.
\begin{thm}
The risk-averse branch MPC with a CVaR cost function can be solved with the following optimization.
\begin{subequations}\label{eq:branch_MPC_risk}
\begin{align}
&\min_{\bfx_i,\bfu_i,\gamma_i,\mu_i^+,\mu_i^-,\sigma_i} \bar{J}_0(\bfx_0,\bfz_0,\bfu_0)+\gamma_0 \nonumber\\
\mathrm{s.t.}& \forall i\in \mathbf{I},\;\forall t=t_i^0+1,...t_i^f, x_{t|b_i}=f(x_{t-1|b_i},u_{t-1|b_i}) \nonumber\\
&\forall i\in\mathbf{I}\backslash \{0\}, x_{t_i^0|b_i}=f(x_{t_i^0-1|\pre(b_i)},u_{t_i^0-1|\pre(b_i)}),\nonumber\\
&\forall i\in\mathbf{I},\; \gamma_i=-\sigma_i+\frac{1}{\alpha}p_{\pi_j|b_i}^\intercal\mu_i^-,\;\mu^+_i\ge 0,\mu^-_i\ge 0, \label{eq:Cvar_nested1}\\
&\forall j\in\ch(b_i),\bar{J}_j(\bfx_j,\bfz_j,\bfu_j)\le -\sigma_i-\mu_{i,j}^++\mu_{i,j}^--\gamma_j.\label{eq:Cvar_nested2}
\end{align}
\end{subequations}
\end{thm}
The derivation of \eqref{eq:branch_MPC_risk} is deferred to the appendix. We omit the linearized branch MPC with risk measure cost as the derivation follows exactly as \eqref{eq:linearized_MPC}.
When $\bar{J}$ is quadratic, \eqref{eq:branch_MPC_risk} is solved as a second order cone programming after linearization.
\section{Branch MPC for highway driving}

This section presents some implementation details of the branch MPC on the highway driving example.
\subsection{Implementation details}
\newsec{Predictive model.} We use a simple predictive model inspired by the backup control barrier function work in \cite{chen2020guaranteed} where the idea is to forward integrate the dynamics following a policy, and check collision along the trajectory. To be specific, for every branching node in the scenario tree, let $x$ and $z$ denote the ego agent and the uncontrolled agent states, respectively. Given a fixed horizon $T$ and a set of policies $\Pi=\{\pi_i\}_{i=1}^m$, we propagate the uncontrolled agent state $z$ under all policies in $\Pi$, and let $\bfz_{\pi_i}$ be the trajectory propagated under $\pi_i$ for the uncontrolled agent. Let $\bfx_{\pi_i}$ denote the planned trajectory in the branch MPC corresponding to the child branch under $\pi_i$, both $\bfx_{\pi_i}$ and $\bfz_{\pi_i}$ last for $T$ steps. Then given safety constraints $\mathcal{C}$ including collision avoidance and lane boundary constraints for the uncontrolled agent, a safety function $h:\mathcal{X}^T\times\mathcal{Z}^T\to\mathbb{R}$ is defined. $h$ is defined such that $h(\bfx,\bfz)\ge 0$ indicates that the uncontrolled agent satisfies $\mathcal{C}$, otherwise it violates $\mathcal{C}$, and the smaller $h$ is, the more $\bfz$ violates $\mathcal{C}$.  the predictive model is simply defined with a softmax function:
\begin{equation}\label{eq:softmax}
  P(\pi_i|x,z)=\frac{\exp(\min\{ (h(\bfx_{\pi_i},\bfz_{\pi_i}),\eta\})}{\sum_j \exp(\min\{h(\bfx_{\pi_j},\bfz_{\pi_j}),\eta\})}.
\end{equation}
where $\eta$ is a parameter that saturates $h$ such that all safe scenarios would have similar probability.

\newsec{Implementation with automatic differentiation.}
To speed up the computation, our implementation heavily relies on automatic differentiation to obtain the gradient. In particular, we use CasADi \cite{Andersson2019} as the computation engine for differentiating the dynamics, the constraints, and the branching probability. In particular, to obtain the branching probability, the scenario tree is propagated forward under $\Pi$ in closed-form and the safety function $h$ is defined with the softmin function instead of the $\min$ function over multiple constraints so that CasADi can perform automatic differentiation. Since we wrote all these functions as compositions of basic functions and lookup tables (which can also be handled by CasADi), the computation for the values and gradients is negligible comparing to the solving time of the SQP. For the branch MPC with expectation cost function in \eqref{eq:branch_MPC}, OSQP \cite{stellato2020osqp} is used to solve the SQP after linearization; for the branch MPC using CVaR cost function in \eqref{eq:branch_MPC_risk}, ECOS \cite{domahidi2013ecos} is called to solve the SOCP. The code can be found on \href{https://github.com/chenyx09/belief-planning}{https://github.com/chenyx09/belief-planning}.

\subsection{Setup for the highway motion planning}
We consider two challenging scenarios in highway driving, overtaking and lane change, and merging.

\newsec{Overtaking and lane change.} Overtaking and lane change is a common task for autonomous vehicles where the ego vehicle needs to overtake the uncontrolled vehicle and perform a lane change to cut in front. A unicycle model is used as the dynamic model for both the ego vehicle and the uncontrolled vehicle:
\begin{equation}\label{eq:unicycle}
  x=\begin{bmatrix}
      X \\
      Y \\
      v \\
      \psi
    \end{bmatrix}\quad \dot{x}=\begin{bmatrix}
                                 v\cos(\psi) \\
                                 v\sin(\psi) \\
                                 a \\
                                 r
                               \end{bmatrix},
\end{equation}
where $X$, $Y$ are the longitudinal and lateral coordinates, $v$ is the velocity, $\psi$ is the heading angle. $a$ and $r$ are the acceleration and yaw rate, which are the control input. The dynamics for the uncontrolled vehicle is exactly the same. The continuous-time dynamics is discretized to obtain the discrete-time model $x^+=f(x,u)$. The cost function is a typical LQR type quadratic cost, i.e.,
\begin{equation*}
  J=(x-x_{ref})^\intercal Q (x-x_{ref})+u^\intercal u,
\end{equation*}
where no penalty is put on $X$. The state constraints are
\begin{equation}\label{eq:constraint}
\begin{aligned}
  &Y_{\min}\le Y\le Y_{\max},\;\psi_{\min}\le\psi\le \psi_{\max},\\
  &\frac{\Delta X e^{\kappa \Delta X}+\Delta Y e^{\kappa \Delta Y}}{ e^{\kappa \Delta X}+ e^{\kappa \Delta Y}}\ge 1,
\end{aligned}
\end{equation}
where $\kappa>0$ is a constant, $\Delta X = \frac{|X-X_z|}{\Delta X_{\max}}$, $\Delta Y = \frac{|Y-Y_z|}{\Delta Y_{\max}}$, $X_z, Y_z$ are the coordinates of the uncontrolled vehicle, $\Delta X_{\max}$ and $\Delta Y_{\max}$ are the longitudinal and lateral clearance. We use the softmax function instead of the $\max$ function so that the expression can be differentiated by automatic differentiation.

The uncontrolled vehicle has three policies, \textit{maintain fixed speed}, \textit{slow down}, and \textit{lane change}, where the direction of \textit{lane change} is towards the ego vehicle as lane change towards the other direction is of no danger to the ego vehicle. During the simulation, the uncontrolled vehicle would update its policy with a fixed frequency, and we tested two update rules: (i) the uncontrolled vehicle randomly sample its policy according to the predictive model in \eqref{eq:predictive_model} (ii) the uncontrolled vehicle always pick the most likely policy, and the simulation results are similar.
\begin{figure}
  \centering
  \includegraphics[width=1\columnwidth]{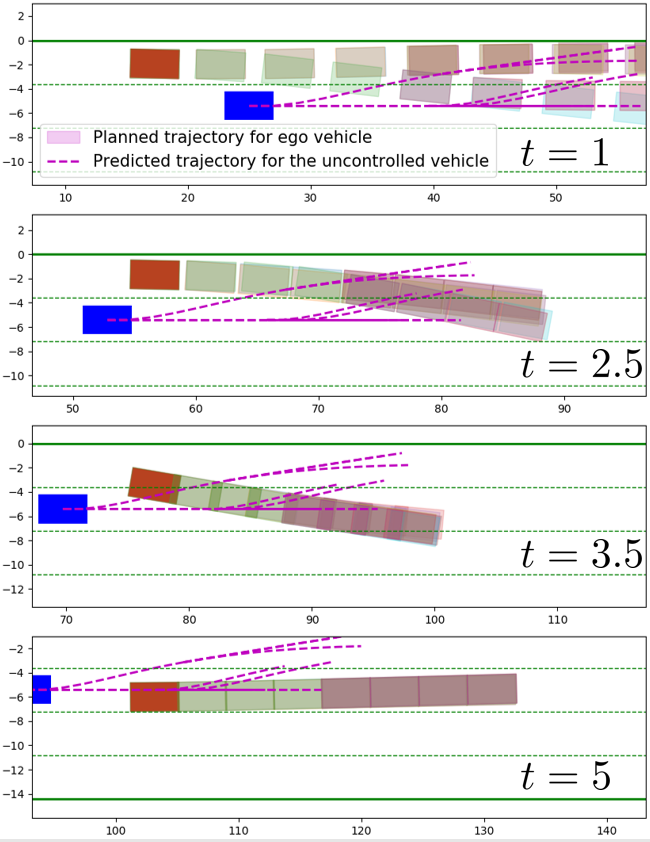}
  \caption{Simulation of the overtaking and lane change under branch MPC}\label{fig:overtaking}
\end{figure}

\begin{figure}[tb]
  \centering
  \includegraphics[width=1\columnwidth]{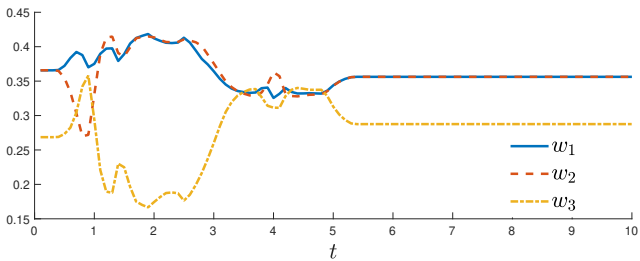}
  \caption{Weights of $b_1$,$b_2$, and $b_3$ over time}\label{fig:weight}
\end{figure}
Fig. \ref{fig:overtaking} shows the snapshots from the simulation of the overtaking and lane change task under the branch MPC controller with the CVaR objective function. $\alpha$ is chosen to be 0.9 ($\text{CVaR}_{0.1}(\cdot)$), which is relatively risk-neutral. As shown in the snapshots, the branch MPC plans a tree of trajectories, corresponding to the multiple possible behaviors in the scenario tree. Fig. \ref{fig:weight} shows the weight changing of $b_1$,$b_2$, and $b_3$ changing over time, which corresponds to the \textit{maintain fixed speed}, \textit{slow down}, and \textit{lane change} policies, respectively (they all have children branches, but their children branches' weights are omitted in the plot). The ego vehicle demonstrates an interesting `nudging' behavior at the beginning since there is a substantial probability that the uncontrolled vehicle may change to the left lane. Since the branch corresponding to the left lane change has a high cost, the gradient of the branching probability in \eqref{eq:linearized_MPC} motivates the ego vehicle to nudge forward so that the weight on $b_3$ reduces. After 2.5 seconds, by nudging forward, the branch MPC determines that the probability of a lane change from the uncontrolled vehicle is very low and eventually it performs the overtaking and cut in front of the uncontrolled vehicle. After the overtaking, the different branches in the trajectory tree converge since all branches of the scenario tree pose little to no danger of collision.

\begin{figure}[tb]
  \centering
  \includegraphics[width=1\columnwidth]{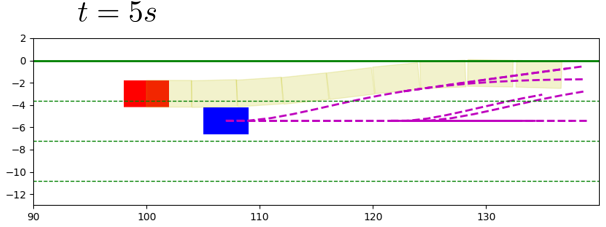}
  \caption{Simulation of the overtaking and lane change under robust MPC}\label{fig:sim_robust}
\end{figure}

As a benchmark, we tested the same task under a robust MPC controller, which only optimizes over one trajectory and tries to avoid all possible trajectories of the uncontrolled vehicle. Due to this limitation, the vehicle is stuck behind the uncontrolled vehicle and is afraid to perform the overtake, as shown in Fig. \ref{fig:sim_robust}.

\begin{figure}[tb]
  \centering
  \includegraphics[width=1\columnwidth]{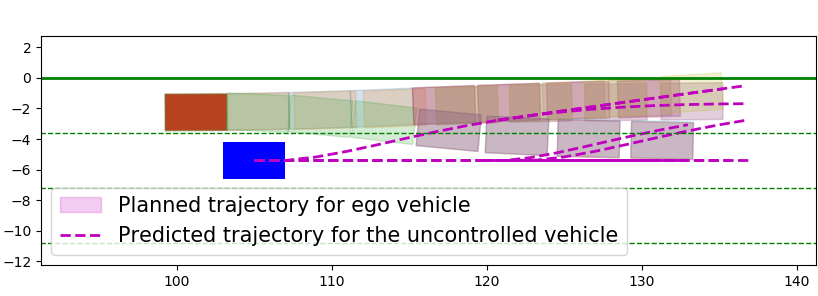}
  \caption{Simulation of the overtaking and lane change under branch MPC with $\text{CVaR}_{0.9}(\cdot)$ cost function}\label{fig:low_alpha}
\end{figure}

To demonstrate the influence of $\alpha$, we run the same simulation with $\alpha=0.1$, which is very risk-averse, and the result is similar to that under the robust MPC, shown in Fig. \ref{fig:low_alpha}. Since the ambiguity set is much larger under the small $\alpha$, the worst-case (maximum) probability of the uncontrolled vehicle performing a lane change to the left is larger, preventing the branch MPC from performing the overtake.

\newsec{Merging.}
The second task considered is merging. The ego vehicle needs to merge into the highway via a ramp and the uncontrolled vehicle drives on the main highway. The cost function is similar to the overtaking task, except that $x_{ref}$ is not a fixed vector, but a linear function. To be specific, the state cost becomes $(Sx-x_{ref})^\intercal Q (Sx-x_{ref})$, which is still quadratic. The merging ramp's geometry is linearized to obtain $S$ and $x_{ref}$ and the cost function is updated at every iteration. The uncontrolled vehicle is equipped with two policies, \textit{maintain fixed speed} and \textit{slow down}.
\begin{figure}[tb]
  \centering
  \includegraphics[width=1\columnwidth]{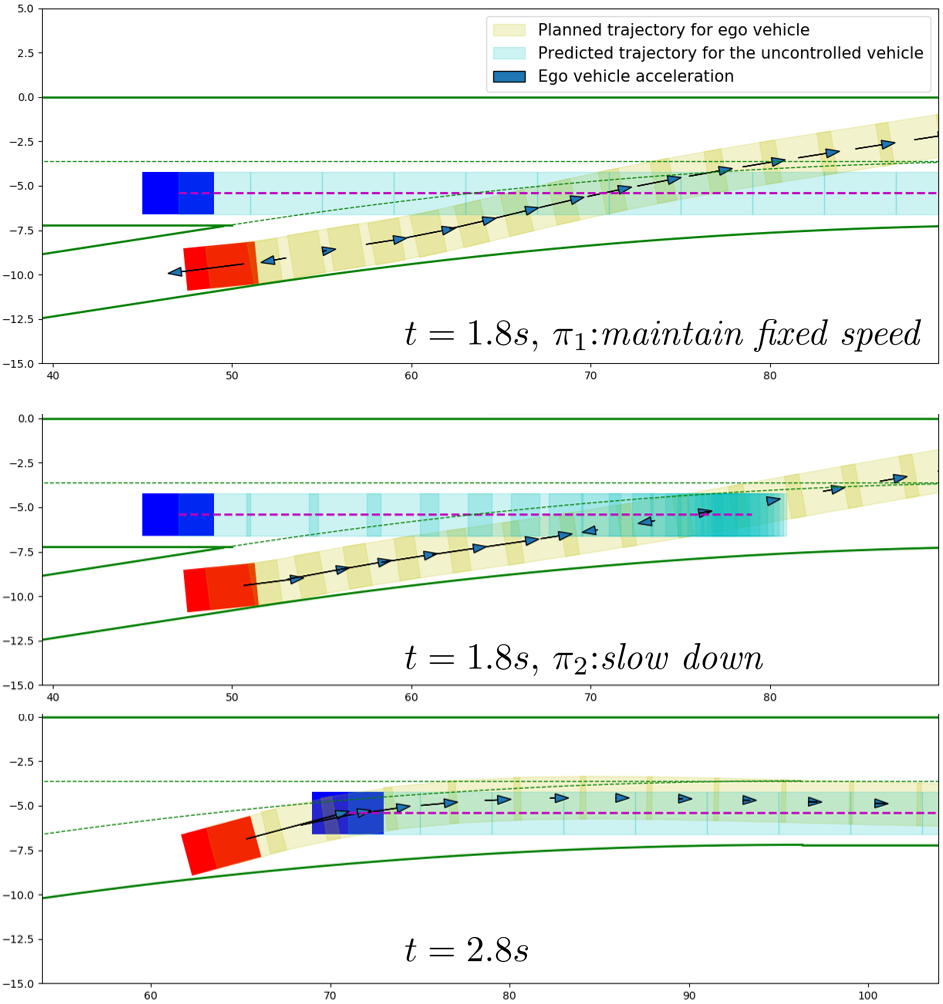}
  \caption{Simulation of the merging task}\label{fig:merging}
\end{figure}

Fig. \ref{fig:merging} shows the snapshots of the simulation. The top and middle plot shows the branches corresponding to $\pi_1$: \textit{maintain fixed speed}, and $\pi_2$: \textit{slow down} at $t=1.8s$. The ego-vehicle prepares to slow down and let the uncontrolled vehicle pass first should the uncontrolled vehicle choose $\pi_1$: and prepares to accelerate and merge in front of the uncontrolled vehicle should the uncontrolled vehicle choose $\pi_2$. Weighing on the probability of the two branches, the ego vehicle eventually chooses to slow down and let the other vehicle pass first.

\newsec{Computation}
In the overtaking and lane change example, $m=3,M=8$, i.e., every branching node has 3 children, branching happens every 8 time steps. The full scenario tree contains 2 layers of branching nodes, with 3 and 9 children, respectively. The ECOS solver is able to solve the branch MPC within 100 ms on an Intel Xeon CPU @ 2.4 GHz. In the merging example, $m=2, M=40$, and the scenario tree only branches once. The solution time is within 80 ms.

\section{Branch MPC for quadruped motion planning}\label{sec:quad}
The branch MPC controller is also tested on a quadruped platform with experiments. In particular, we consider two quadruped robots, one remotely controlled (uncontrolled robot), the other under the branch MPC controller (ego robot). the operator would give waypoints for the ego robot to reach, and the ego robot would navigate itself to the waypoints while interacting with the uncontrolled robot.

\subsection{Implementation details}
\newsec{Platform setup}
A Unitree A1 quadruped was used as the ego robot, with locomotion performed via an inverse dynamics-based trotting controller built off the work in \cite{buchli2009inverse}. This controller was executed via an off-board i7-8565U CPU @ 1.80GHz with 16GB RAM, which computed desired joint torques, velocities, and positions and communicated with the A1 motor drivers over UDP at 1kHz. The trot controller tracked body-frame forward, lateral, and angular velocity commands from the branch MPC controller using the motion primitive framework in \cite{ubellacker2021motion_primitives}.
The uncontrolled robot was a teleoperated Vision 60 from Ghost Robotics using the proprietary walking controller for locomotion.
To get global location of the two quadrupeds, an OptiTrack motion capture system was used with 6 cameras and 5 markers per robot. The position and orientation data was streamed over a ROS node at 125 Hz to the branch MPC controller on the off-board computer.

\newsec{Dynamic model and setup for branch MPC}
The actual quadruped model is a 36 dimensional hybrid highly nonlinear model, which is not suitable for the interactive motion planning. We use a modified unicycle model on the high-level planning layer and sends command to the low-level controller which tracks the desired motion. In particular, the following unicycle model is used:
\begin{equation}\label{eq:quad_unicycle}
    x=\begin{bmatrix}
      X \\
      Y \\
      \psi
    \end{bmatrix}\quad \dot{x}=\begin{bmatrix}
                                 v_x\cos(\psi)-v_y \sin(\psi) \\
                                 v_x\sin(\psi) + v_y \cos(\psi) \\
                                 r
                               \end{bmatrix},
\end{equation}
where $X,Y,\psi$ are the global coordinates and heading angle of the robot, $v_x,v_y$, and $r$ are the longitudinal velocity, lateral velocity, and yaw rate, which are the inputs of the high-level branch MPC controller. Compared to the vehicle case, the quadruped robots have much smaller masses and thus can accelerate and decelerate much faster. Therefore, we directly use velocity as input and rely on the low-level controller to track the desired velocity. Lateral velocity is added as an input since the quadruped has the ability to sidestep, yet it incurs a much larger cost as it is not the desired motion.

\begin{figure}[tb]
  \centering
  \includegraphics[width=1\columnwidth]{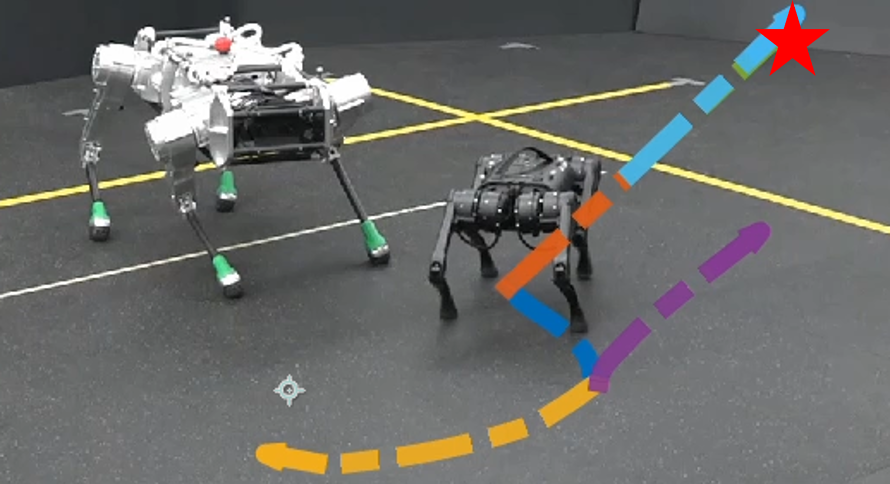}
  \includegraphics[width=1\columnwidth]{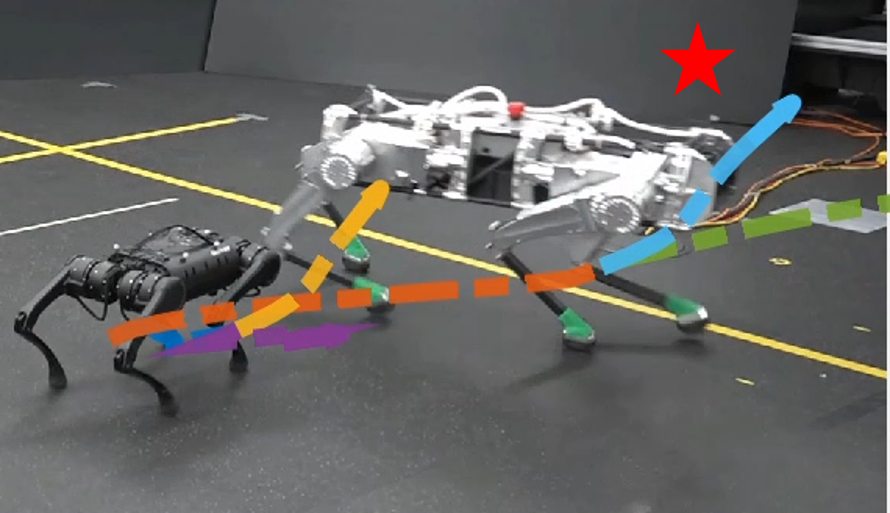}
  \caption{Snapshots of the quadruped experiment}\label{fig:quad_experiment}
\end{figure}

The uncontrolled quadruped is remotely controlled by a human operator. The branch MPC equips the uncontrolled robot with two policies: moving forward with constant speed and stopping, corresponding to the two operation modes, not yielding and yielding. The video of the experiment can be found \href{https://youtu.be/W3jzoMjAZsQ}{here}. Fig. \ref{fig:quad_experiment} shows two snapshots of the experiment where the star shows the position of the waypoint, and the trajectory tree is plotted with one color for each branch. In both snapshots, the ego robot plans different trajectories in preparation for the uncontrolled robot to either keep moving forward, or to stop. In the top figure, the ego robot would keep moving straight to the goal should the uncontrolled robot stop, and side step to the right should the uncontrolled robot move forward. In the bottom figure, the ego robot would wait until the uncontrolled robot moves away and cross from behind should the uncontrolled robot move forward, and cross in front of the uncontrolled robot should the uncontrolled robot stop.

\begin{figure}[tb]
    \centering
    \includegraphics[width=1\columnwidth]{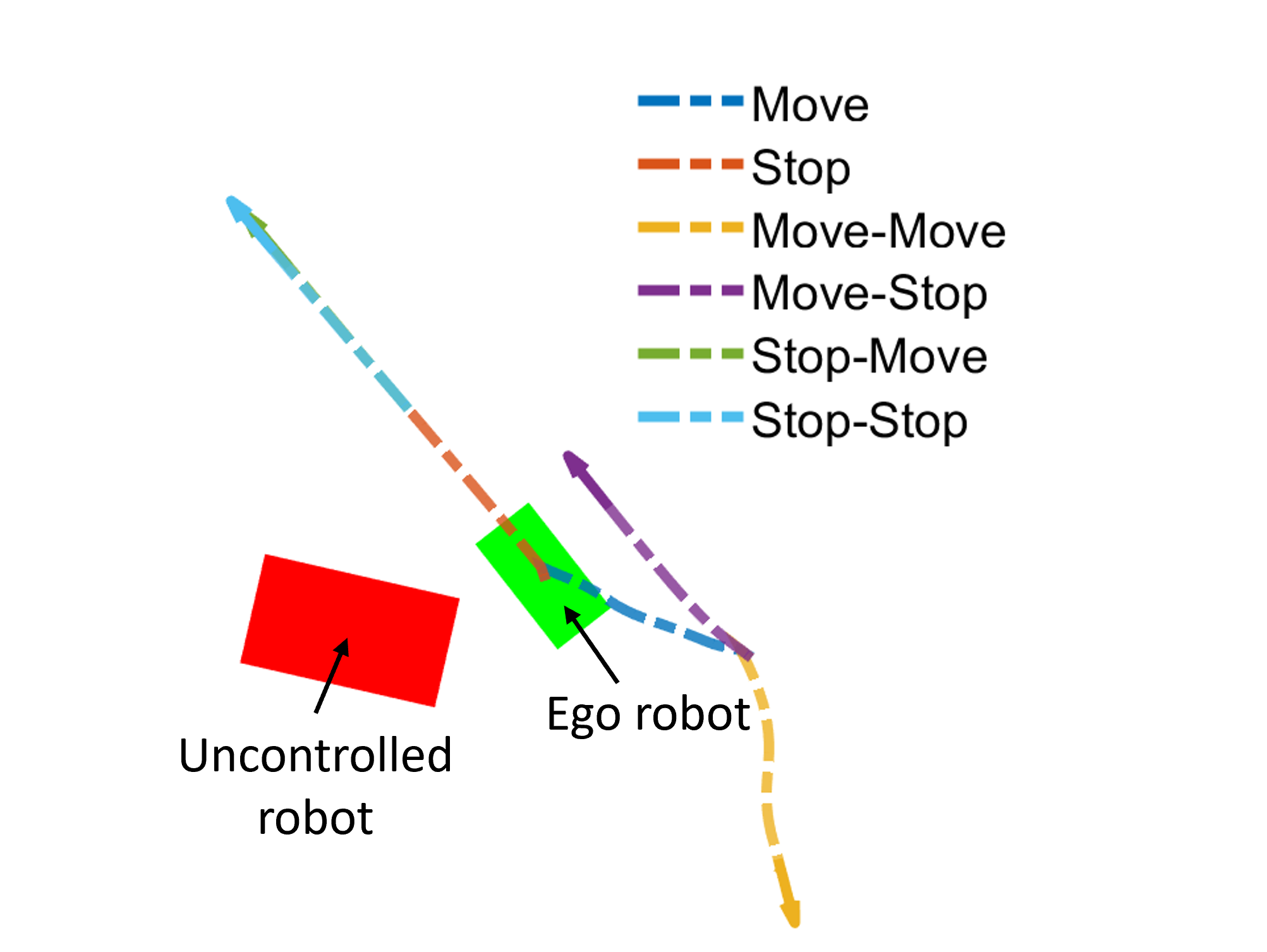}
    \includegraphics[width=1\columnwidth]{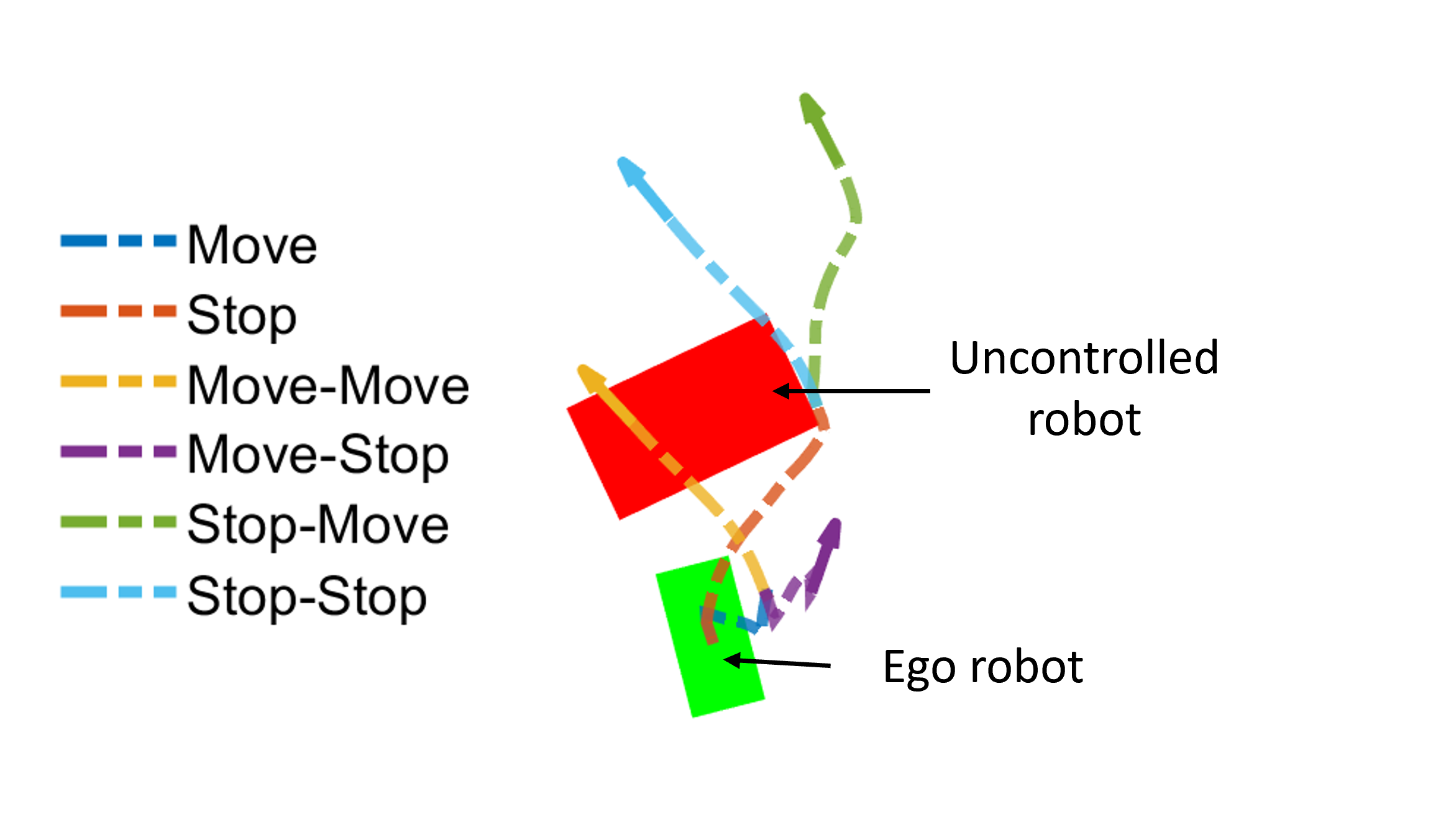}
    \caption{Top-down view of the motion plan}\label{fig:top_down_quad}
\end{figure}

Fig. \ref{fig:top_down_quad} shows the same two moments in Fig. \ref{fig:quad_experiment} from top down. The dashed lines represent the branches corresponding to the assumed motion of the uncontrolled vehicle. For instance, in the first plot, the blue dashed line shows the branch in the trajectory tree corresponding to the uncontrolled agent moving forward, in which case the ego robot would move back to avoid collision; the orange branch shows the plan should the uncontrolled robot stop, and the ego robot would move past the uncontrolled robot towards the waypoint. Both branches have two children branches corresponding to the subsequent motion of the uncontrolled robot.

\newsec{Stability and optimality of the SQP}
The nonlinear, nonconvex MPC was solved by linearizing around the solution from the last time step and formulate a quadratic approximation of the problem. The sequential quadratic programming strategy may cause stability issues as the linearization point may change drastically between time steps. This issue was not observed in the highway driving example as the vehicle's heading angle was constrained to a narrow range; yet we observed instability of the SQP in the quadruped experiment. We were able to maintain stability after adding a proximal cost, i.e., $J_{\text{prox}}=||\bfx-\hat{\bfx}||$, which penalizes the deviation from the linearization point. With stability ensured by proximal terms, the SQP may still get stuck in local minimums, resulting in compromised performance. In future works, we plan to combine the branch MPC with sampling based motion planning tools such as probabilistic roadmaps and use the latter to provide linearization point to the SQP solver to prevent the local minimum problem.

\section{Conclusion and future works}
We present a branch Model Predictive Control framework for reactive motion planning for autonomous agents in the presence of uncontrolled agents. The branch MPC approximates the reactive behavior of the uncontrolled agent with a finite set of policies to build a scenario tree, where each branch has an associated probability determined by the reactive model. A feedback policy is then solved in the form of a trajectory tree, whose topology is determined by the scenario tree. Furthermore, risk measures such as the Conditional Value at Risk (CVaR) are used to tune the tradeoff between performance and robustness. We demonstrate the efficacy of the method on a highway driving example for autonomous vehicles and a quadruped motion planning example. The result shows that the branch MPC is able to capture the reactivity of the uncontrolled agent in future prediction steps and plan human-like behaviors for the autonomous agent that balances liveness and safety.

There are several aspects of the proposed algorithm that can be improved. First, the timing of branching is assumed to be fixed (branch in every $M$ steps), which may not be realistic. We would like to extend the current framework to event-triggered asynchronous branching, which poses additional challenge to the SQP, yet is closer to the real reactive agents. Second, the current framework does not take previous observations into account, which can be incorporated by introducing beliefs and an observation model into the MPC. Third, the reactive model is assumed to be given: we plan to use data-driven methods to provide a more rigorous and justifiable pipeline to obtain the reactive model. Furthermore, the current framework would struggle with multiple uncontrolled agents due to the exponential complexity of the product policy space. Therefore, we plan to develop a data-driven algorithm that takes the interaction between uncontrolled agents into account to learn reactive models that output the joint behavior of all uncontrolled agents, which could be multimodal but not as the product of individual behaviors. By directly generating the joint behavior, we can eliminate the unrealistic combination of uncontrolled agents' behavior and greatly reduce the complexity of the branch MPC computation.

\appendix
\section{Derivation of the DRO for CVaR optimization}\label{sec:DRO}
We shall only look at the formulation for CVaR optimization at one branching node: the overall CVaR branch MPC in \eqref{eq:branch_MPC_risk} is formulated by composing several CVaR constraints. As mentioned above, $\phi_i$ denotes the cost of subsequent branches, and let $p=[p_{j_1},...p_{j_m}]$ denote the probability distribution over the children branches where $\ch(b_i)=\{j_1,...,j_m\}$. Let $\gamma_i=\text{CVaR}_{1-\alpha}(\phi_i)$. It follows then that $\phi_i=\bar{J}_{j}+\gamma_j$ with probability $p_j$ for $j\in\ch(b_i)$. For the children branches that end with a leaf node, $\gamma_j=0$. Using the dual form of CVaR,
\begin{equation*}
  \gamma_i = \text{CVaR}_{1-\alpha}(\phi_i)=\max_{q\in\mathscr{A}_i} \sum_j q_j (\bar{J}_j+\gamma_j),
\end{equation*}
where $\mathscr{A}_i=\{q=[q_{j_1},...q_{j_m}]\in\mathbb{R}^m|q\ge 0 ,\sum_j q_j=1,q_j\le \frac{1}{\alpha}p_j,j\in\ch(b_i)\}$. By Lagrange duality, the Lagrangian of the maximization is
\begin{equation*}
  \mathcal{L} = \sum_j q_j (\bar{J}_j+\gamma_j) + (\mu^+_i)^\intercal q + (\mu^-_i)^\intercal(\frac{1}{\alpha} p-q) + \sigma_i (\sum_j q_j-1),
\end{equation*}
where $\mu^+_i=[\mu^+_{i,j_1},...,\mu^+_{i,j_m}]\ge 0$ is the Lagrange multiplier for the constraint $q\ge 0$, $\mu^-_i=[\mu^-_{i,j_1},\mu^-_{i,j_m}]\ge 0$ is the Lagrange multiplier for the constraint $q\le \frac{1}{\alpha} p$, and $\sigma_i$ is the Lagrange multiplier for $\sum_j q_j=1$. It is easy to see that the feasible set has a nonempty interior. Therefore, strong duality holds:
\[\max_{q\in\mathscr{A}_i}\sum_j q_j(\bar{J}_j+\gamma_j)=\min_{\mu_i^+\ge 0,\mu_i^-\ge 0,\sigma_i}\mathcal{L}.\]
The dual problem is then
\begin{equation*}
\begin{aligned}
\min_{\mu_i^+\ge 0,\mu_i^-\ge 0,\sigma_i}& -\sigma_i+\frac{1}{\alpha}p^\intercal \mu_i^-\\
\mathrm{s.t.}\;\;& \bar{J}_j(\bfx_j,\bfz_j,\bfu_j)\le -\sigma_i-\mu_{i,j}^++\mu_{i,j}^--\gamma_j.
\end{aligned}
\end{equation*}
Note that by strong duality, the $\gamma_i$ is equal to the objective of the dual problem, i.e.,
\begin{equation*}
\gamma_i= -\sigma_i+\frac{1}{\alpha}p^\intercal \mu_i^-,
\end{equation*}
The constrained minimization for the dual problem is then equivalent to the constraints in \eqref{eq:Cvar_nested1} and \eqref{eq:Cvar_nested2}.

\balance
\bibliographystyle{myieeetran}
\bibliography{mybib}
\end{document}